\documentclass[twocolumn,conference,ruled,linesnumbered]{IEEEtran}
\usepackage[T1]{fontenc}
\usepackage[latin9]{inputenc}
\usepackage{color}
\usepackage{array}
\usepackage{algorithm2e}
\usepackage{amsmath}
\usepackage{amssymb}
\usepackage{graphicx}

\makeatletter

\providecommand{\tabularnewline}{\\}

\usepackage{verbatim}
\usepackage{amsthm}

\theoremstyle{plain}
\theoremstyle{plain}
\theoremstyle{plain}
\providecommand{\factname}{Fact}
\theoremstyle{remark}

\newcommand{\subparagraph}{}
\usepackage[]{titlesec}

\usepackage{amsfonts}
\usepackage{cite}
\usepackage{array}

\usepackage{subfigure}

\DeclareMathOperator{\logdet}{logdet}
\DeclareMathOperator{\tr}{tr}
\DeclareMathOperator{\diag}{diag}

\DeclareMathOperator*{\minimize}{minimize}
\DeclareMathOperator*{\maximize}{maximize}
\DeclareMathOperator*{\subjectto}{subject~to}
\DeclareMathOperator*{\nulls}{null}

\providecommand{\lemmaname}{Lemma}
\providecommand{\theoremname}{Theorem}
\usepackage{tikz}
\usepackage{pgfplots}

\usepackage{tabularx}
\usetikzlibrary{spy,arrows}
\usepackage{subfigure}

\usetikzlibrary{plotmarks,positioning}
\usetikzlibrary{spy,shapes,patterns,trees}
\usepackage{array}
\usetikzlibrary{backgrounds}
\tikzset{mark options={line width=1pt,solid}, font=\footnotesize}%

\usepackage{tikz}
\usepackage{pgf}\usepackage{pgfplots}\usetikzlibrary{automata,positioning}
\usetikzlibrary{arrows,backgrounds,plotmarks}
\pgfplotsset{plot coordinates/math parser=false}

\makeatother

\begin{document}
\title{On Estimating Maximum Sum Rate of MIMO Systems with Successive Zero-Forcing
Dirty Paper Coding and Per-antenna Power Constraint}
\author{\IEEEauthorblockN{Thuy M. Pham$^{\ast}$, Ronan Farrell$^{\ast}$, and Le-Nam Tran$^{\ddagger}$}\IEEEauthorblockA{$^{\ast}$Department of Electronic Engineering, Maynooth University,
Ireland. Email: \{minhthuy.pham, ronan.farrell\}@mu.ie\\
$^{\ddagger}$School of Electrical and Electronic Engineering, University
College Dublin, Ireland. Email: nam.tran@ucd.ie}}
\maketitle
\begin{abstract}
In this paper, we study the sum rate maximization for successive zero-forcing
dirty-paper coding (SZFDPC) with per-antenna power constraint (PAPC).
Although SZFDPC is a low-complexity alternative to the optimal dirty
paper coding (DPC), efficient algorithms to compute its sum rate are
still open problems especially under practical PAPC. The existing
solution to the considered problem is computationally inefficient
due to employing high-complexity interior-point method. In this study,
we propose two new low-complexity approaches to this important problem.
More specifically, the first algorithm achieves the optimal solution
by transforming the original problem in the broadcast channel into
an equivalent problem in the multiple access channel, then the resulting
problem is solved by alternating optimization together with successive
convex approximation. We also derive a suboptimal solution based on
machine learning to which simple linear regressions are applicable.
The approaches are analyzed and validated extensively to demonstrate
their superiors over the existing approach.
\end{abstract}


\section{Introduction}

The preliminary studies on multiple-input multiple-output (MIMO) capacity
showed that single-user MIMO capacity can be achieved by Gaussian
input signaling \cite{Telatar:MIMO:1999,Foschini:MIMO:1998}. In \cite{Weingarten06p3936},
Weingarten \textit{et al.} further proved that the entire capacity
region of a Gaussian MIMO broadcast channel (BC) is achievable through
optimal dirty paper coding (DPC) method. In reality, the optimal precoding
method is not only high-complexity but also difficult to implement.
As a result, there are growing interests in suboptimal techniques
such as zero-forcing (ZF) and successive zero-forcing dirty-paper
coding (SZFDPC) \cite{Caire:ZFDPC:2003,Dabbagh:SZFDPC:2007,Spencer:BD:2004,Thuy17:ZF:PAPC,Nam:SZFDPC:2012,Nam:beamdesign:ZFDPC:2012}.

Traditionally, the majority of the research on MIMO capacity assumes
a sum power constraint (SPC) for which efficient algorithms can be
derived \cite{Telatar:MIMO:1999,Foschini:MIMO:1998,Yu:IterativeWF:MAC:2004,Jindal:IterativeWF:BC:2005,Liu08p3664}.
Nevertheless, solutions to SPC problems may result in nonlinear distortions
if the allocated power is beyond the power limit of one or several
power amplifiers. Towards this end, per-antenna power constraint (PAPC)
is more realistic and of particular interest\cite{MaiVu:MIMO:PAPC:2011,Yu:TransmitOptimization:PerAntennaConstraint:2007,Thuy:SU:PAPC,Thuy:MU:PAPC}
.

To the best of authors' knowledge, only Tran \textit{et al.} characterized
the achievable rate region of SZFDPC under PAPC by customized interior-point
method \cite{Nam:beamdesign:ZFDPC:2012,Nam:SZFDPC:2012}. However,
this second-order-based method is not attractive to large-scale MIMO
systems due to high complexity. In this study, we propose two algorithms
to obtain sum rates of MIMO systems under PAPC and SZFDPC. The first
algorithm achieves the optimal solution by alternating optimization
(AO) while the second one exploits machine learning (ML) to arrive
at a suboptimal one. In particular, our contributions include the
following:
\begin{itemize}
\item A novel AO-based algorithm is proposed to obtain the optimal solution.
Specifically, the original maximization in the BC is transformed into
an equivalent minimax problem in the multiple access channel (MAC),
then an efficient iterative algorithm is derived based on successive
convex approximation (SCA).
\item In case the optimal approach is feasible but inefficient, an ML-based
approach, which trades off the complexity and the optimal solution,
is a good alternative. In fact, our ML approach relies on linear regression
and thus is appealing to applications such as massive MIMO.
\item We report for the first time the comparison of SZFDPC and others precoding
methods under PAPC. Moreover, our ML-based approach can be applicable
to similar capacity-related problems.
\end{itemize}
The remainder of the paper is organized as follows. A description
of the system model is in Section \ref{sec:model}. The methods of
computing the sum rate for SZFDPC are described in Section \ref{sec:algorithm}
followed by the numerical results in Section \ref{sec:results}. Finally,
we conclude the paper in Section \ref{sec:conclusion}.

\emph{Notation}: Standard notations are used in this paper. Bold lower
and upper case letters represent vectors and matrices, respectively.
$\mathbf{I}$ defines an identity matrix, of which the size can be
easily inferred from the context; $\mathbb{C}^{M\times N}$ denotes
the space of $M\times N$ complex matrices; $\mathbf{H}^{\dagger}$
and $\mathbf{H}^{T}$ are Hermitian and normal transpose of $\mathbf{H}$,
respectively; $\mathbf{H}_{i,j}$ is the $(i,j)$th entry of $\mathbf{H}$;
$|\mathbf{H}|$ is the determinant of $\mathbf{H}$; $\nulls(\mathbf{H})$
stands for a basis of the null space of $\mathbf{H}$; $|\mathbf{x}|$
denotes the absolute value of $\mathbf{x}$; $\diag(\mathbf{H})$,
where $\mathbf{H}$ is a square matrix, returns the vector of diagonal
elements of $\mathbf{H}$. Furthermore, we denote the Euclidean norm
by $||\cdot||$ and $[x]_{+}=\max(x,0)$.

\section{System Model\label{sec:model}}

Consider a MIMO BC consisting of a base station (BS) and $K$ users.
The BS and each user $k$ are equipped with $N$ and $M$ antennas
respectively. The channel matrix for user $k$ is denoted by $\mathbf{H}_{k}\in\mathbb{C}^{{M}\times{N}}$.
Normally, a user suffers interference from all other users in the
system. For user $k$ in the SZFDPC scheme, the interference caused
by users $j<k$ is cancelled by DPC, while that caused by users $j>k$
is nulled out by zero-forcing technique. In this way, a MIMO BC can
be decomposed into parallel interference-free channels. We refer the
interested reader to \cite{Nam:SZFDPC:2012} and references therein
for a more detailed description of the SZFDPC scheme.

The sum rate of SZFDPC can be characterized through solving the sum
rate (SRMax) problem under PAPC which is formulated as \begin{subequations}\label{eq:MUMIMO:SR:SZFDPC:PAPC:Primal}
\begin{align}
{\textstyle \underset{\{\mathbf{S}_{k}\succeq\mathbf{0}\}}{\maximize}} & \quad{\textstyle \sum_{k=1}^{K}\log|\mathbf{I}+\mathbf{H}_{k}\mathbf{S}_{k}\mathbf{H}_{k}^{\dagger}|}\\
\subjectto & \quad{\textstyle \mathbf{H}_{j}\mathbf{S}_{k}\mathbf{H}_{j}^{\dagger}=0,~\forall j<k\label{eq:SZFDPC:zerointference}}\\
 & \quad{\textstyle \sum_{k=1}^{K}[\mathbf{S}_{k}]_{i,i}\leq P_{i},~i=1,2,\ldots,N}
\end{align}
\end{subequations} where $\mathbf{S}_{k}\succeq\mathbf{0}$ is the
input covariance matrix for user $k$. The constraint in \eqref{eq:SZFDPC:zerointference}
is imposed to suppress the interference from users $j<k$ as mentioned
above.

Due to the use of zero-forcing method, SZFDPC is a suboptimal transmission
strategy compared to DPC. However, SZFDPC does not cancel multiuser
inference only by zero-forcing technique since DPC is still invoked
for this purpose. Thus, SZFDPC can achieve a performance close to
that of DPC, which was reported in various previous studies \cite{Nam:GreedyScheduling:SZFDPC:2010,Nam:SZFDPC:2012,Nam:beamdesign:ZFDPC:2012}.
We note that for SZFDPC (i.e. \eqref{eq:MUMIMO:SR:SZFDPC:PAPC:Primal})
to be feasible, it should hold that $N>(K-1)M$ which is assumed in
this paper. This dimension condition basically imposes a constraint
on the maximum number of users that can be supported simultaneously.
When the number of demanding users increases, a user scheduling algorithm
is required and this problem was studied in \cite{Nam:GreedyScheduling:SZFDPC:2010}
where several efficient user selection methods were proposed for SZFDPC.
We also remark that the interference cancelling process is performed
sequentially after each user, and thus user ordering in SZFDPC is
important. Optimal user ordering requires solving a combinatorial
optimization problem but efficient user order algorithms were also
proposed in \cite{Nam:GreedyScheduling:SZFDPC:2010}. In this paper
we simply assume the natural user ordering for SZFDPC and focus on
the precoder design.

In order to simplify the formulation in \eqref{eq:MUMIMO:SR:SZFDPC:PAPC:Primal},
let $\breve{\mathbf{H}}_{k}=[\mathbf{H}_{1}^{\dagger},\mathbf{H}_{2}^{\dagger},\ldots\mathbf{H}_{k-1}^{\dagger}]^{\dagger}$,
$\breve{\mathbf{V}}_{k}=\nulls(\breve{\mathbf{H}})$, and $\dot{\mathbf{H}}_{k}=\mathbf{H}_{k}\mathbf{\breve{V}}_{k}$.
Intuitively, $\dot{\mathbf{H}}_{k}$ is called the effective channel
of user $k$. The optimal $\mathbf{S}_{k}$ in \eqref{eq:MUMIMO:SR:SZFDPC:PAPC:Primal}
is then given by $\mathbf{S}_{k}=\breve{\mathbf{V}}_{k}\dot{\mathbf{S}}_{k}\breve{\mathbf{V}}_{k}^{\dagger}$,
where $\dot{\mathbf{S}}_{k}$ is the optimal solution to the following
problem
\begin{equation}
\begin{array}{rl}
\underset{\{\dot{\mathbf{S}}_{k}\succeq\mathbf{0}\}}{\maximize} & \sum_{k=1}^{K}\log|\mathbf{I}+\dot{\mathbf{H}}_{k}\dot{\mathbf{S}}_{k}\dot{\mathbf{H}}_{k}^{\dagger}|\\
\subjectto & \sum_{k=1}^{K}[\mathbf{\breve{V}}_{k}\dot{\mathbf{S}}_{k}\mathbf{\breve{V}}_{k}^{\dagger}]_{i,i}\leq P_{i},\:\forall i.
\end{array}\label{eq:SZFDPC:primal}
\end{equation}

\section{Proposed Algorithms\label{sec:algorithm}}

\subsection{Alternating Optimization}

Inspired by the work in \cite{Thuy17:ZF:PAPC}, we extend the AO approach
to our considered problem. More specifically, by extending Theorem
2 of \cite{Nam:beamdesign:ZFDPC:2012}, we can show that \eqref{eq:SZFDPC:primal}
can be equivalently transformed into the following minimax problem
in the dual MAC
\begin{equation}
\begin{array}{rl}
\underset{\mathbf{Q}\succeq\mathbf{0}}{\min}\ \underset{\{\bar{\mathbf{S}}_{k}\succeq\mathbf{0}\}}{\max} & \sum_{k=1}^{K}\log\frac{|\mathbf{\breve{V}}_{k}^{\dagger}\mathbf{Q}\mathbf{\breve{V}}_{k}+\dot{\mathbf{H}}_{k}^{\dagger}\bar{\mathbf{S}}_{k}\dot{\mathbf{H}}_{k}|}{|\mathbf{\breve{V}}_{k}^{\dagger}\mathbf{Q}\mathbf{\breve{V}}_{k}|}\\
\subjectto & \sum_{k=1}^{K}\tr(\bar{\mathbf{S}}_{k})=P\\
 & \tr(\mathbf{Q}\mathbf{P})=P,\mathbf{Q}:\mathsf{diagonal}.
\end{array}\label{eq:SZFDPC:dual}
\end{equation}
The relationship between optimal solutions of \eqref{eq:SZFDPC:primal}
and \eqref{eq:SZFDPC:dual} is given by 
\begin{equation}
\dot{\mathbf{S}}_{k}=(\mathbf{\breve{V}}_{k}^{\dagger}\mathbf{Q}\mathbf{\breve{V}}_{k})^{-1/2}\mathbf{U}\mathbf{V}^{\dagger}\bar{\mathbf{S}}_{k}\mathbf{V}\mathbf{U}^{\dagger}(\mathbf{\breve{V}}_{{\color{black}k}}^{\dagger}\mathbf{Q}\mathbf{\breve{V}}_{{\color{black}k}})^{-1/2}
\end{equation}
 where $\mathbf{U},\mathbf{V}$ are obtained from the singular value
decomposition of $(\mathbf{\breve{V}}_{{\color{black}k}}^{\dagger}\mathbf{Q}\mathbf{\breve{V}}_{{\color{black}k}})^{-1/2}\dot{\mathbf{H}}_{{\color{black}k}}^{\dagger}$
\cite{Vishwanath:duality_achievable:2003}. In light of AO algorithm
in \cite{Thuy17:ZF:PAPC,Thuy:SU:PAPC,Thuy:MU:PAPC}, we first fix
$\mathbf{Q}$ and consider the following problem 
\begin{equation}
\begin{array}{rl}
{\maximize} & \sum_{k=1}^{K}\log|\mathbf{\breve{V}}_{k}^{\dagger}\mathbf{Q}^{n}\breve{\mathbf{V}}_{k}+\dot{\mathbf{H}}_{k}^{\dagger}\bar{\mathbf{S}}_{k}\dot{\mathbf{H}}_{k}|\\
\subjectto & \sum_{k=1}^{K}\tr(\bar{\mathbf{S}}_{k})=P;\{\bar{\mathbf{S}}_{k}\succeq\mathbf{0}\}.
\end{array}\label{eq:SZFDPC:dual:findS}
\end{equation}
Problem \eqref{eq:SZFDPC:dual:findS} is the one of finding the capacity
of parallel interference-free MIMO channels under a sum power constraint,
which can be solved efficiently by the classical water-filling algorithm.

We now consider the problem of finding $\mathbf{Q}$ for given {\color{black}$\{\bar{\mathbf{S}}_{k}^{n}\}$}.
To this end, we apply the following $\logdet$ inequality: 
\begin{multline}
\log|\mathbf{\breve{V}}_{k}^{\dagger}\mathbf{Q}\breve{\mathbf{V}}_{k}+\dot{\mathbf{H}}_{k}^{\dagger}{\color{black}\bar{\mathbf{S}}_{k}^{n}}\dot{\mathbf{H}}_{k}|\leq\log|\boldsymbol{\Phi}_{k}^{n}|+\\
\tr\Bigl(\mathbf{\breve{V}}_{k}\boldsymbol{\Phi}_{k}^{-n}\mathbf{\breve{V}}_{k}^{\dagger}\bigl(\mathbf{Q}-\mathbf{Q}^{n}\bigr)\Bigr)\label{eq:MUMIMO:SR:SZFDPC-1storder}
\end{multline}
where $\boldsymbol{\Phi}_{k}^{n}\triangleq\mathbf{\breve{V}}_{k}^{\dagger}\mathbf{Q}^{n}\breve{\mathbf{V}}_{k}+\dot{\mathbf{H}}_{k}^{\dagger}{\color{black}\bar{\mathbf{S}}_{k}^{n}}\dot{\mathbf{H}}_{k}$,
and $\boldsymbol{\Phi}_{k}^{-n}$ stands for $\bigl(\boldsymbol{\Phi}_{k}^{n}\bigr)^{-1}$.
In the $(n+1)$th iteration of the proposed algorithm to solve \eqref{eq:SZFDPC:dual},
$\mathbf{Q}^{n+1}$ is the solution to the following problem 
\begin{equation}
\begin{array}{rl}
\min & \hspace{-5pt}\sum_{k=1}^{K}\bigl(\tr\bigl(\mathbf{\breve{V}}_{k}\boldsymbol{\Phi}_{k}^{-n}\mathbf{\breve{V}}_{k}^{\dagger}\mathbf{Q}\bigr)-\log|\mathbf{\breve{V}}_{k}^{\dagger}\mathbf{Q}\breve{\mathbf{V}}_{k}|\bigr)\\
\mathrm{s.t.} & \hspace{-5pt}\tr(\mathbf{Q}\mathbf{P})=P,\mathbf{Q}:\mathsf{diagonal};\mathbf{Q}\succeq\mathbf{0}.
\end{array}\label{eq:SR:SZFDPC:Q-min}
\end{equation}
Since $\mathbf{Q}$ is diagonal, \eqref{eq:SR:SZFDPC:Q-min} indeed
reduces to 
\begin{equation}
\begin{array}{rl}
{\min} & ~\boldsymbol{\alpha}^{T}\mathbf{q}-\sum_{k=1}^{K}\log|\mathbf{\breve{V}}_{k}^{\dagger}\diag(\mathbf{q})\breve{\mathbf{V}}_{k}|\\
\mathrm{s.t.} & ~\mathbf{p}^{T}{\mathbf{q}}=P
\end{array}\label{eq:SR:SZFDPC:Q-min:compact}
\end{equation}

where $\boldsymbol{\alpha}=\sum_{k=1}^{K}\bigl(\diag(\mathbf{\breve{V}}_{k}\boldsymbol{\Phi}_{k}^{-n}\mathbf{\breve{V}}_{k}^{\dagger})\bigr)$.

It's worth noting that the feasible set of \eqref{eq:SR:SZFDPC:Q-min:compact}
is in fact a simplex. As shown shortly, projection onto a simplex
can be done efficiently by closed-form expressions and this motivates
us to solve \eqref{eq:SR:SZFDPC:Q-min:compact} by a gradient projection
(GP) method, which is outlined in Algorithm \ref{alg:SR:SZFDPC:findingQ}. 

\begin{algorithm}
\global\long\def\thealgorithm{\arabic{algorithm}}%
 \caption{\color{black} The Proposed GP Algorithm for Solving \eqref{eq:SR:SZFDPC:Q-min:compact}.
\label{alg:SR:SZFDPC:findingQ}}

\setcounter{AlgoLine}{0}
\SetAlgoNoLine\SetNlSty{textnormal}{}{}

\KwIn{$\mathbf{p}$ , $\epsilon>0$ }

Initialization: $\tau=1+\epsilon$, $m=0$, $\mathbf{q}_{0}=\mathbf{1}_{N}^{T}$.

\While{$\tau>\epsilon$} { Calculate the gradient $\tilde{\mathbf{g}}_{m}=\nabla f(\mathbf{q}_{m})=\boldsymbol{\alpha}-\sum_{k=1}^{K}\diag(\mathbf{\breve{V}}_{k}^{\dagger}(\mathbf{\breve{V}}_{k}\diag(\mathbf{q}_{m})\mathbf{\breve{V}}_{k}^{\dagger})^{-1}\mathbf{\breve{V}}_{k})$.\label{alg:SR:SZFDPC:findingQ:grad}

Choose an appropriate positive scalar $s_{m}$ and create $\tilde{\mathbf{q}}_{m}=\mathbf{q}_{m}-s_{m}\tilde{\mathbf{g}}_{m}$.

Project $\tilde{\mathbf{q}}_{m}$ onto $\mathcal{Q}_{q}=\{\mathbf{p}^{T}\mathbf{q}=P,\mathbf{q}\geq\mathbf{0}\}$
to obtain $\bar{\mathbf{q}}_{m}$.

Choose appropriate stepsize $\beta_{m}$ and set $\mathbf{q}_{m+1}=\mathbf{q}_{m}+\beta_{m}(\bar{\mathbf{q}}_{m}-\mathbf{q}_{m})$
using the Armijo rule \cite{Armijo66p1}.\label{linesearch}

$\tau=|\nabla f(\mathbf{q}_{m})^{T}(\mathbf{q}_{m+1}-\mathbf{q}_{m})|$.

$m:=m+1$. } \KwOut{$\mathbf{q}^{m}$ as the optimal solution to
\eqref{eq:SR:SZFDPC:Q-min:compact}.}
\end{algorithm}


In Algorithm \ref{alg:SR:SZFDPC:findingQ} the subscript denotes the
iteration index and $\tilde{\mathbf{g}}_{m}$ is the gradient of the
objective at iteration $m$ computed as in line \ref{alg:SR:SZFDPC:findingQ:grad}.
Projection of $\tilde{\mathbf{q}}_{m}$ onto the feasible set of \eqref{eq:SR:SZFDPC:Q-min:compact}
is equivalent to solving the following problem 
\begin{equation}
\begin{array}{rl}
{\minimize} & ~\frac{1}{2}||{\mathbf{q}}-\tilde{\mathbf{q}}_{m}||^{2}\\
\subjectto & ~\mathbf{p}^{T}{\mathbf{q}}=P;{\mathbf{q}}\geq{0}.
\end{array}\label{eq:szfdpc-projection}
\end{equation}
This optimization problem can be solved efficiently by a water-filling-like
algorithm. Specifically, the partial Lagrangian function of \eqref{eq:szfdpc-projection}
is written as 
\begin{equation}
\mathcal{L}(\mathbf{q},\boldsymbol{\gamma})=\frac{1}{2}||{\mathbf{q}}-\tilde{\mathbf{q}}_{m}||^{2}+\gamma(\mathbf{p}^{T}{\mathbf{q}}-P).
\end{equation}
For a given $\gamma$, it is easy to see that the optimal solution
to $\underset{\mathbf{q}\geq0}{\max}\ \mathcal{L}(\mathbf{q},\boldsymbol{\gamma})$
is given by ${\mathbf{q}}^{*}=[\tilde{\mathbf{q}}_{m}-\gamma\mathbf{p}]_{+}$.
The optimal $\gamma$ such that $\mathbf{p}^{T}{\mathbf{q}}^{*}=P$
can be simply found by bisection method. Note that when the PAPC is
the same for all antennas, i.e., $p_{i}=P/N$, for all $i=1,2,\ldots,N$,
the feasible set becomes a canonical simplex for which more efficient
algorithms for projection are available \cite{Condat16p575}. The
overall algorithm to solve the SRMax problem with SZFDPC and PAPC
is summarized in Algorithm \ref{alg:SR:SZFDPC:alg2}. The convergence
of Algorithm \ref{alg:SR:SZFDPC:alg2} can be proved similarly to
\cite[Appendix B]{Thuy17:ZF:PAPC} and thus is skipped for the sake
of brevity. 

\begin{algorithm}
\global\long\def\thealgorithm{\arabic{algorithm}}%
 \caption{\color{black} Proposed Algorithm for SRMax Problem with SZFDPC based
on AO. \label{alg:SR:SZFDPC:alg2}}

\setcounter{AlgoLine}{0}
\SetAlgoNoLine\SetNlSty{textnormal}{}{}

\KwIn{$\mathbf{Q}:=\boldsymbol{\mathbf{Q}}^{0}$ diagonal matrix
of positive elements, $\epsilon>0$}

Initialization: Set $n:=0$ and $\tau=1+\epsilon$, $\breve{\mathbf{H}}_{1}={\mathbf{H}}_{1}$,
and $\mathbf{\breve{V}}$$_{1}=\mathbf{I}$. For each $k\geq2$, create
$\breve{\mathbf{H}}_{k}=[\mathbf{H}_{1}^{\dagger},\mathbf{H}_{2}^{\dagger},\ldots\mathbf{H}_{k-1}^{\dagger}]^{\dagger}$,
$\mathbf{\breve{V}}_{k}=\nulls(\breve{\mathbf{H}}_{k})$, and $\dot{\mathbf{H}}_{k}=\mathbf{H}_{k}\mathbf{\breve{V}}_{k}$.

\While{$\tau>\epsilon$}{

Apply the water-filling algorithm to solve \eqref{eq:SZFDPC:dual:findS}.
Denote the optimal solution by {\color{black}$\{\bar{\mathbf{S}}_{k}^{n}\}$}.

If $n\geq1$, let $\tau=|f^{\text{SZF-DPC}}(\mathbf{Q}^{n},\{\bar{\mathbf{S}}_{k}^{n}\})-f^{\text{SZF-DPC}}(\mathbf{Q}^{n-1},\{\bar{\mathbf{S}}_{k}^{n-1}\})|$
where $f^{\text{SZF-DPC}}(.)$ denotes the objective in \eqref{eq:SZFDPC:dual}.

For each $k$, set $\boldsymbol{\Phi}_{k}^{n}=(\mathbf{\breve{V}}_{k}^{\dagger}\mathbf{Q}^{n}\breve{\mathbf{V}}_{k}+\dot{\mathbf{H}}_{k}^{\dagger}{\color{black}\bar{\mathbf{S}}_{k}^{n}}\dot{\mathbf{H}}_{k})$.

Find $\mathbf{Q}^{n+1}$ using Algorithm \ref{alg:SR:SZFDPC:findingQ}.

$n:=n+1$.}

\KwOut{ $\{\bar{\mathbf{S}}_{k}^{n}\}_{k=1}^{K}$ and apply the
BC-MAC transformation to compute optimal $\{\dot{\mathbf{S}}_{k}^{n}\}_{k=1}^{K}$.}
\end{algorithm}

\subsection{A Feature Design-based Approach}

Following similar arguments in \cite{Thuy17:ZF:PAPC}, the interior-point-based
approach to solve the considered problem has the complexity up to
$\mathcal{O}(K^{3}N^{6})$ while the total per-iteration complexity
of Algorithm \ref{alg:SR:SZFDPC:alg2} is {\color{black}$\mathcal{O}(KN^{3})$}
flops. On the one hand, Algorithm \ref{alg:SR:SZFDPC:alg2} dominates
the existing approach and reduces the complexity significantly, but
on the other hand, it still experiences high complexity in case of
massive MIMO settings where $\frac{N}{K}\geq10$. In such cases, we
can employ the following ML approach to obtain a suboptimal solution
since this approach can adapt quickly to any changes in the systems
while retaining the satisfactory performance. 

Assuming that we execute Algorithm \ref{alg:SR:SZFDPC:alg2} to generate
optimal sum rates $\mathbf{y}$ based on $\mathbf{X}=[\mathbf{x}_{1},\mathbf{x}_{2},\ldots,\mathbf{x}_{s}]\in\mathbb{C}^{{p}\times{s}}$
inputs where $s$ is the number of samples. Note that $\mathbf{x}_{i}$
contain $p$ features of the power constraints and channel coefficients.
If we simply apply arbitrary ML algorithms, the errors will be extremely
prohibitive due to the fact that the considered problem is nonlinear
in nature with respect to either power constraint or the channel matrix
(c.f. Fig. \ref{fig:MUMIMO:SR:ML:Estimate}). On the other hand, nonlinear
ML algorithms are much more difficult to investigate since there are
no available solutions to this type of optimization. Even the optimal
solution mentioned above already contains many nonlinear terms. Here,
we propose a novel two-step preprocessing method to transform the
inputs into another feature space so that linear regression algorithms
are applicable. Herein, we will refer to this approach as feature
design (FD) based approach.  %
\begin{tabular}{|>{\raggedright}p{0.95\columnwidth}|}
\hline 
\vspace{-1pt}
\textbf{Step 1}: Select a set of features $\check{\mathbf{x}}$ by
customizing the principle component analysis (PCA)-based algorithm
in \cite{PCA:selection}:

- Choose the number of eigenvectors whose eigenvalues are larger than
1 .

- Select the features based on $l$ largest contribution

\textbf{Step 2}: Transform $\check{\mathbf{x}}$ into higher feature
space by $\boldsymbol{\phi}(\check{\mathbf{x}})=[1,\log_{b}(|\check{\mathbf{x}}|)]^{T}$.

\vspace{3pt}
\tabularnewline
\hline 
\end{tabular}

Note that instead of choosing a number of largest eigenvalues of the
covariance matrix randomly\cite{PCA:selection}, we choose $d$ eigenvalues
which are larger than 1. As a result, we can form a new matrix $\tilde{\mathbf{U}}=[\tilde{\mathbf{u}}_{1},\tilde{\mathbf{u}}_{2},\ldots,\tilde{\mathbf{u}}_{d}]$
corresponding to those eigenvalues. To select the most dominant features,
we calculate the contribution measure 
\begin{equation}
\vartheta_{i}=\sum_{j=1}^{d}|\tilde{u}_{i,j}|
\end{equation}
 where $i=1,2,\ldots,p$. Then we select the desired features with
respect to $l$ largest contribution $\vartheta_{i}$. Again, we avoid
random selection of $l$ whose appropriate value is not easy to justify
in practice. Instead, we propose to choose $l$ based on the matrix
size and the number of users: 
\begin{equation}
l=N+Kr\label{eq:Con:criteria}
\end{equation}
 where $N$ and $K$ are the number of transmit antennas and the number
of users, respectively; $r=\min(M,N)$ where $M$ is the number of
receive antennas. Note that $l<p$ from \eqref{eq:Con:criteria} and
we can therefore obtain a new matrix with reduced dimension $\check{\mathbf{X}}=[\mathbf{\check{\mathbf{x}}}_{1},\mathbf{\check{\mathbf{x}}}_{2},\ldots,\mathbf{\check{\mathbf{x}}}_{s}]\in\mathbb{C}^{{l}\times{s}}$. 

In fact there are no criteria to choose a function to transform the
inputs into another space where efficient algorithms can be derived.
In our approach, we rely on the characteristics of the problem to
propose a transform function. Specifically, recall that the considered
sum rate is a $\logdet$ function, thus we can transform these features
into new space features where linear model are possible using the
following
\begin{equation}
\boldsymbol{\phi}(\check{\mathbf{x}})=\left[\begin{array}{c}
1\\
\log_{b}(|\check{\mathbf{x}}|)
\end{array}\right]\label{eq:Feature:trans}
\end{equation}
where $b$ is the base of the logarithm. Under this assumption, an
output is given by
\begin{equation}
y_{i}\approx\boldsymbol{\phi}(\check{\mathbf{x}}_{i})^{T}\hat{\mathbf{w}}.
\end{equation}
As a result of this formulation, we can apply any linear regression
algorithms such as ordinary least square (OLS), ridge regression or
principal component regression (PCR)\cite[Chapter 6]{Theodoridis:2015:MLB:2792418}
to find an appropriate estimator. In the numerical results, we will
show the effectiveness of our proposed approach in comparison with
other algorithms which do not take the feature design into account.

\section{Numerical Results\label{sec:results}}

In this section, we numerically evaluate the performance of the proposed
algorithms. For all iterative algorithms of comparison, we set an
error tolerance of $\epsilon=10^{-6}$ as the stopping criterion.
The initial value $\mathbf{Q}^{0}$ in the corresponding proposed
algorithm is set to the identity matrix for all simulations and the
total transmit power is simply set to $0$ dBW, if not mentioned otherwise.
The number of receive antennas is fixed to $M=2$ and and the power
limit for all antennas is equal to $P/N$. Other simulation parameters
are specified for each setup. The codes are executed on a 64-bit desktop
that supports 8 Gbyte RAM and Intel CORE i7. 

In the first experiment, we compare the average sum rate of different
precoding methods i.e., SZFDPC, ZF\cite{Thuy17:ZF:PAPC}, and DPC\cite{Thuy:MU:PAPC}
with PAPC over a large number of channel realizations. Under large-scale
MIMO setup ($N/K\ge10$), three methods obtain the identical value
since the correlation between channels approaches to zero. However,
under normal MIMO settings, there is still a big gap between the capacity
for ZF and that of SZFDPC and DPC. In general, SZFDPC always achieves
a near-capacity rate. 

\begin{figure}
\centering \includegraphics[bb=66bp 307bp 287bp 478bp,scale=0.7]{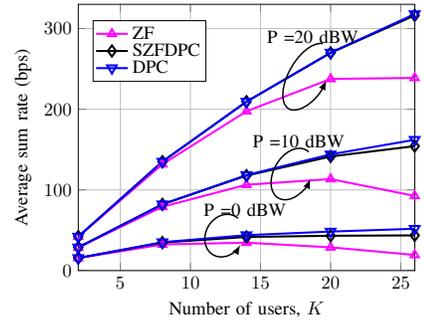}
\caption{Average sum rate versus the number of users for MIMO systems with
PAPC, $N=64$ transmit antennas.}
\label{fig:MUMIMO:Avgsum}
\end{figure}

\begin{figure}[h]
\centering \includegraphics[bb=60bp 313bp 288bp 474bp,scale=0.7]{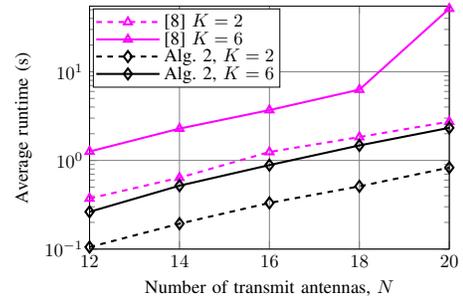}
\caption{Runtime versus the number of transmit antennas for solving the sum
rate maximization problem with SZFDPC scheme. Two methods are compared:
Algorithm \ref{alg:SR:SZFDPC:alg2}, and the interior-point method
in \cite{Nam:SZFDPC:2012}.}
\label{fig:MUMIMO:Runtime:SZFDPC}
\end{figure}

Since Algorithm \ref{alg:SR:SZFDPC:alg2} and other benchmark scheme
of consideration all generate the optimal solution to the corresponding
problem, we mainly compare their complexity. In particular, we compare
the runtime of Algorithm \ref{alg:SR:SZFDPC:alg2} with the interior-point
method proposed in \cite{Nam:SZFDPC:2012}. Fig. \ref{fig:MUMIMO:Runtime:SZFDPC}
plots the average runtime as a function of the number of transmit
antennas $N$ for finding the maximum sum rate of SZFDPC. We observe
that Algorithm \ref{alg:SR:SZFDPC:alg2} performs water-filling and
GP to find $\bar{\mathbf{S}},\mathbf{Q}$, which results in lower
computation time compared with \cite{Nam:SZFDPC:2012} which uses
the barrier interior-point method. In general, the barrier method
and other second-order optimization methods are known to achieve a
superlinear rate but its per-iteration cost increases quickly with
the problem size. This is actually consistent with what is shown in
Fig. \ref{fig:MUMIMO:Runtime:SZFDPC} for the barrier-method \cite{Nam:SZFDPC:2012}.

As can be seen from Fig. \ref{fig:MUMIMO:Runtime:SZFDPC}, our AO-based
algorithm, though has low complexity, may take a few seconds or more
to execute when we increase the problem size. Thus, a method of lower
complexity which can strike a balance between the sum rate and complexity
is of interest. In the following, we will investigate the performance
of our ML-based approach to such scenarios. The PAPC ratio is chosen
randomly, whereas SNR is chosen from the set $\textrm{SNR}=[0,10,20,30,40]$
dBW. For each MIMO setting we generated 240 samples. Also, we simply
use natural log to transform the feature space in \eqref{eq:Feature:trans}.

\begin{figure}[h]
\centering \includegraphics[bb=63bp 297bp 285bp 485bp,scale=0.7]{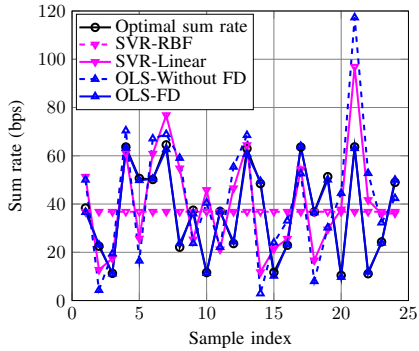}
\caption{Sum rate estimate of linear and nonlinear regressions with and without
feature design approach, $N=32$ transmit antennas, $M=2$ receive
antennas and $K=2$ users.}
\label{fig:MUMIMO:SR:ML:Estimate}
\end{figure}

In Fig. \ref{fig:MUMIMO:SR:ML:Estimate}, we compare the optimal and
estimate sum rates of a MIMO system with linear and nonlinear regression
methods. More specifically, we utilize support vector regression (SVR)
with radial basis function (RBF) kernel \cite{Smola2004} for nonlinear
regression. Here, we train on 216 samples and test on 24 samples.
As can be seen from the figure, without proper preprocessing methods
both OLS and SVR fail to fit the data due to nonlinear nature of the
problem. However, the results of the simple OLS, which take feature
design into account, are very close to optimal solutions. The performance
has also proved the feasibility of our approach. 

In the last experiment, we consider the effectiveness of our approach
in terms of average relative root mean square error (aRRMSE) \cite{LI2013139}
over large samples with varying number of transmit and receive antennas.
In particular, we obtain the aRRMSE by executing 10-fold cross-validation
using three simple linear ML algorithms: OLS, Ridge and PCR. According
to \cite{LI2013139}, a learning model is considered good and excellent
when $10\%<\textrm{aRRMSE}<20\%$ and $\textrm{aRRMSE}<10\%$, respectively.
Interestingly, the ML-based method shows a sufficiently low error
rate, especially when $\frac{N}{K}\geq10$. From our observations,
the training matrices are invertible and the eigenvalues are larger
than 1, thus the performance of OLS and PCR is the same and has minor
difference in comparison with that of ridge regression. Unsurprisingly,
this observation coincides with the properties of these regression
methods. 

\begin{figure}[h]
\centering \includegraphics[bb=65bp 326bp 284bp 460bp,scale=0.7]{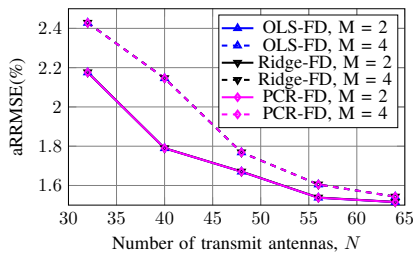}
\caption{aRRMSE of OLS, Ridge regression and PCR with feature design and $K=2$
users.}
\label{fig:MUMIMO:SR:ML:aRRMSE}
\end{figure}

\section{Conclusions}

\label{sec:conclusion} We have proposed two low-complexity methods
to compute sum rates of MIMO systems under PAPC and SZFDPC. The experiments
using the optimal approach have stated that the SZFDPC can obtain
near-capacity rates whereas the ZF scheme still operates far from
the optimal capacity boundary for a specified number of users. The
suboptimal ML-based method is more advantageous in case of large-scale
MIMO settings. Extensive numerical results have demonstrated the superiority
of the proposed algorithms over the existing interior-point method.
More importantly, our ML-based approach can be applicable to a class
of similar problems.

\end{document}